\newcommand\apjcls{1}
\newcommand\aastexcls{2}
\newcommand\othercls{3}

\newcommand\papercls{\aastexcls}
\documentclass[tighten, times, twocolumn]{aastex62}

\if\papercls \apjcls
\usepackage{apjfonts}
\else\if\papercls \othercls
\usepackage{epsfig}
\usepackage{margin}
\usepackage{times}
\fi\fi
\usepackage{ifthen}
\usepackage{natbib}
\usepackage{amssymb, amsmath}
\usepackage{appendix}
\usepackage{etoolbox}
\usepackage[T1]{fontenc}
\usepackage{paralist}
\usepackage{mathrsfs}
\usepackage{multirow}
\usepackage{rotating} 
\usepackage[graphicx]{realboxes}
\usepackage{bm}
\usepackage{appendix}
\usepackage[utf8]{inputenc}

\if\papercls \apjcls
\newcommand\aas{\ref@jnl{AAS Meeting Abstracts}}
\newcommand\dps{\ref@jnl{AAS/DPS Meeting Abstracts}}
\newcommand\maps{\ref@jnl{MAPS}}
\else\if\papercls \othercls
\usepackage{astjnlabbrev-jh}
\fi\fi

\if\papercls \aastexcls
\hypersetup{citecolor=blue, 
            linkcolor=blue, 
            menucolor=blue, 
            urlcolor=blue}  
\else
\usepackage[
bookmarks=true,           
bookmarksnumbered=true,   
colorlinks=true,          
citecolor=blue,           
linkcolor=blue,           
menucolor=blue,           
urlcolor=blue,            
linkbordercolor={0 0 1},  
pdfborder={0 0 1},
frenchlinks=true]{hyperref}
\fi

\if\papercls \othercls

\else

\fi

\providecommand{\adsurl}[1]{\href{#1}{ADS}}

\makeatletter
\patchcmd{\NAT@citex}
  {\@citea\NAT@hyper@{%
     \NAT@nmfmt{\NAT@nm}%
     \hyper@natlinkbreak{\NAT@aysep\NAT@spacechar}{\@citeb\@extra@b@citeb}%
     \NAT@date}}
  {\@citea\NAT@nmfmt{\NAT@nm}%
  \NAT@aysep\NAT@spacechar\NAT@hyper@{\NAT@date}}{}{}

\patchcmd{\NAT@citex}
  {\@citea\NAT@hyper@{%
     \NAT@nmfmt{\NAT@nm}%
     \hyper@natlinkbreak{\NAT@spacechar\NAT@@open\if*#1*\else#1\NAT@spacechar\fi}%
      {\@citeb\@extra@b@citeb}%
     \NAT@date}}
  {\@citea\NAT@nmfmt{\NAT@nm}%
  \NAT@spacechar\NAT@@open\if*#1*\else#1\NAT@spacechar\fi\NAT@hyper@{\NAT@date}}
  {}{}
\makeatother

\makeatletter
\DeclareRobustCommand{\lowcase}[1]{\@lowcase#1\@nil}
\def\@lowcase#1\@nil{\if\relax#1\relax\else\MakeLowercase{#1}\fi}
\makeatother

\DeclareSymbolFont{UPM}{U}{eur}{m}{n}
\DeclareMathSymbol{\umu}{0}{UPM}{"16}
\let\oldumu=\umu
\renewcommand\umu{\ifmmode\oldumu\else\math{\oldumu}\fi}

\if\papercls \othercls

\else

\fi

\let\oldsim=\sim
\renewcommand\sim{\ifmmode\oldsim\else\math{\oldsim}\fi}
\let\oldpm=\pm
\renewcommand\pm{\ifmmode\oldpm\else\math{\oldpm}\fi}
\newcommand\by{\ifmmode\times\else\math{\times}\fi}

\newbox{\wdbox}
\renewcommand\c{\setbox\wdbox=\hbox{,}\hspace{\wd\wdbox}}
\renewcommand\i{\setbox\wdbox=\hbox{i}\hspace{\wd\wdbox}}

\newcount\timect
\newcount\hourct
\newcount\minct
\newcommand\now{\timect=\time \divide\timect by 60
         \hourct=\timect \multiply\hourct by 60
         \minct=\time \advance\minct by -\hourct
         \number\timect:\ifnum \minct < 10 0\fi\number\minct}

\catcode`@=11

\newcommand\comment[1]{}

\newcommand\commenton{\catcode`\%=14}

\renewcommand\math[1]{$#1$}
\newcommand\mathshifton{\catcode`\$=3}

\let\atab=&
\newcommand\atabon{\catcode`\&=4}

\let\oldmsp=\sp
\let\oldmsb=\sb
\def\sp#1{\ifmmode
           \oldmsp{#1}%
         \else\strut\raise.85ex\hbox{\scriptsize #1}\fi}
\def\sb#1{\ifmmode
           \oldmsb{#1}%
         \else\strut\raise-.54ex\hbox{\scriptsize #1}\fi}
\newbox\@sp
\newbox\@sb
\def\sbp#1#2{\ifmmode%
           \oldmsb{#1}\oldmsp{#2}%
         \else
           \setbox\@sb=\hbox{\sb{#1}}%
           \setbox\@sp=\hbox{\sp{#2}}%
           \rlap{\copy\@sb}\copy\@sp
           \ifdim \wd\@sb >\wd\@sp
             \hskip -\wd\@sp \hskip \wd\@sb
           \fi
        \fi}
\def\msp#1{\ifmmode
           \oldmsp{#1}
         \else \math{\oldmsp{#1}}\fi}
\def\msb#1{\ifmmode
           \oldmsb{#1}
         \else \math{\oldmsb{#1}}\fi}

\def\supon{\catcode`\^=7}

\def\subon{\catcode`\_=8}

\def\supsubon{\supon \subon}

\newcommand\actcharon{\catcode`\~=13}

\newcommand\paramon{\catcode`\#=6}

\comment{And now to turn us totally on and off...}

\newcommand\reservedcharson{ \commenton  \mathshifton  \atabon  \supsubon 
                             \actcharon  \paramon}

\catcode`@=12
\reservedcharson

\if\papercls \apjcls

\else

\fi

\newcommand\chisq{\ifmmode{\chi\sp{2}}\else\math{\chi\sp{2}}\fi}
\newcommand\redchisq{\ifmmode{ \chi\sp{2}\sb{\rm red}}
                    \else\math{\chi\sp{2}\sb{\rm red}}\fi}
\newcommand\Teq{\ifmmode{T\sb{\rm eq}}\else$T$\sb{eq}\fi}
\newcommand\mjup{\ifmmode{M\sb{\rm Jup}}\else$M$\sb{Jup}\fi}
\newcommand\rjup{\ifmmode{R\sb{\rm Jup}}\else$R$\sb{Jup}\fi}
\newcommand\msun{\ifmmode{M\sb{\odot}}\else$M\sb{\odot}$\fi}
\newcommand\rsun{\ifmmode{R\sb{\odot}}\else$R\sb{\odot}$\fi}
\newcommand\mearth{\ifmmode{M\sb{\oplus}}\else$M\sb{\oplus}$\fi}
\newcommand\rearth{\ifmmode{R\sb{\oplus}}\else$R\sb{\oplus}$\fi}

\usepackage{soul}
\setulcolor{blue}
\setstcolor{red}
\sethlcolor{green}

\usepackage{cleveref}

\def\kmsMpc{\ensuremath\,\rm{km}\,\rm{s}^{-1}/\rm{Mpc}}

\makeatletter
\patchcmd\H@refstepcounter{\protected@edef}{\protected@xdef}{}{}
\makeatother

\pdfoutput=1
\usepackage{enumitem}
\usepackage{mathrsfs}
\usepackage{cleveref}
\defcitealias{Lemos:2018smw}{L18}
\defcitealias{Riess_2018}{R18}
\shorttitle{$H_0$ Reconstruction with SN Ia, BAO and GLTD}
\shortauthors{Lyu {\em et al.}}

\begin{document}

\title{$H_0$ Reconstruction with Type Ia Supernovae, Baryon Acoustic Oscillation and Gravitational Lensing Time-Delay}

\author{Meng-Zhen~Lyu}
\correspondingauthor{Meng-Zhen~Lyu}
\affiliation{Department of Astronomy, Beijing Normal University, Beijing 100875, China}

\author{Balakrishna S. Haridasu}
\affiliation{Dipartimento di Fisica, Universit\`a di Roma "Tor Vergata", Via della Ricerca Scientifica 1, I-00133, Roma, Italy} 
\affiliation{Sezione INFN, Universit\`a di Roma ”Tor Vergata”, Via della Ricerca Scientifica 1, I-00133, Roma, Italy}

\author{Matteo Viel}
\affiliation{SISSA-International School for Advanced Studies, Via Bonomea 265, 34136 Trieste, Italy}
\affiliation{INFN, Sezione di Trieste, Via Valerio 2, I-34127 Trieste, Italy}
\affiliation{INAF - Osservatorio Astronomico di Trieste, Via G. B. Tiepolo 11, I-34143 Trieste, Italy}
\affiliation{IFPU, Institute for Fundamental Physics of the Universe, via Beirut 2, 34151 Trieste, Italy}

\author{Jun-Qing~Xia}
\affiliation{Department of Astronomy, Beijing Normal University, Beijing 100875, China}

\email{lvmz@mail.bnu.edu.cn \\ haridasu@roma2.infn.it \\ viel@sissa.it \\ xiajq@bnu.edu.cn}

\begin{abstract}
There is a persistent $H_0$-tension, now at more than $\gtrsim 4\sigma$ level, between the local distance ladder value and the \emph{Planck} cosmic microwave background measurement, in the context of flat $\Lambda$CDM model. We reconstruct $H(z)$ in a cosmological-model-independent way using three low-redshift distance probes including the latest data from baryon acoustic oscillation, Type Ia supernova and four gravitational lensing Time-Delay observations. We adopt general parametric models of $H(z)$ and assume a Gaussian prior on the sound horizon at drag epoch, $r_{\mathrm s}$, from \emph{Planck} measurement. The reconstructed $H_0$ using Pantheon SN Ia and BAO data are consistent with the \emph{Planck} flat $\Lambda$CDM value. When including the GLTD data, $H_0$ increases mildly, yet remaining discrepant with the local measurement at $\sim 2.5\sigma$ level. Our reconstructions being blind to the dark sectors at low redshift, we reaffirm the earlier claims that the Hubble tension is not likely to be solved by modifying the energy budget of the low-redshift universe. We further forecast the constraining ability of future realistic mock BAO data from DESI and GLTD data from LSST, combining which, we anticipate that the uncertainty of the inferred $H_0$ would be improved by $\sim 38\%$, reaching $\sigma_{H_0} \approx 0.56$ uncertainty level. 
\end{abstract}


\section{Introduction}
\label{set:intro}

The flat $\Lambda$CDM model is a remarkably successful cosmological model. It describes well many observational results, especially at large scales, including the cosmic microwave background (CMB) radiation, light element abundance as the relic of Big Bang nucleosynthesis, galaxy clustering, Lyman-$\alpha$ forest observations and also low redshift distance probes. However, there exists a strong tension for the present Hubble expansion rate ($H_0$), between the direct measurement using distance ladder of local Universe \citep{Riess:2016jrr,Riess:2019cxk,Yuan:2019npk}, and the \emph{Planck} estimate \citep{Ade:2015xua, Aghanim:2018eyx} from CMB within the context of $\Lambda$CDM \citep{Bernal16,Verde:2019ivm,Raveri:2018wln}. One important aspect is that the discordance, since the first release of \emph{Planck} data \citep{Ade:2013zuv}, has become even more prominent due to the improved precision of both these measurements which are at $\sim 9\%$ difference, now reaching a significance of $\gtrsim 4\sigma$ \citep{Riess:2019cxk}. More recent low-redshift gravitational lensing time-delay measurements, independent of the local distance ladder, also a the tension at high significance \cite{Wong19}.  The $H_0$ tension, persisting and severely increasing, indicates that it should not merely be regarded as a statistical fluctuation, and is more likely to point to a failure of the standard $\Lambda$CDM model, as also noted in \cite{Verde:2019ivm}, or due to unknown systematics in the data.

CMB provides a stringent constraint on $H_0$ by combining the measurements of angular location and relative height of the acoustic oscillation of the baryon-photon fluid frozen at last scattering surface at $z\approx1100$. However, the measurement is model-dependent and influenced by possible extensions to the $\Lambda$CDM model, such as the dark energy equation of state parameter $w$\footnote{One possible way to relieve the Hubble tension is allowing phantom dark energy \citep{Vagnozzi19,DiValentino:2017zyq}. This might however have discrepancy with the low redshift BAO measurements, which constraints better the $w \lesssim -1$ range, see e.g., \cite{Bernal16, Aubourg_2015, Haridasu17a, Park19}.} or the curvature $\Omega_{\rm k}$, which as is well-known further aggravates the tension. Thus modifying either the early or the local Universe physics can, in principle, alter the $H_0$ constraints from CMB measurements.

Modification to the $\Lambda$CDM model often involves ingredients beyond the standard physics, although the existence of dark matter and dark energy within the $\Lambda$CDM framework has already established the necessity for "new" physics. Preferable approaches can be to modified dark energy model and different gravitational field behavior \citep{DiValentino:2017rcr,Qing-Guo:2016ykt,DiValentino:2017zyq,Zhao:2017urm,Poulin:2018zxs,Choi:2019jck,Banihashemi:2018has,Khosravi:2017hfi, Umilt__2015, Rossi_2019, Ballardini_2016}, such as an early dark energy \citep{xiaviel09,Karwal:2016vyq,Poulin:2018cxd,Mortsell:2018mfj,Ye20}, interaction between dark sectors \citep{Ko:2016uft,PhysRevD.96.103501,PhysRevD.97.043513,Archidiacono:2019wdp}, interacting dark energy model \citep{DiValentino:2017iww, Yang:2018euj, PhysRevD.88.063501,PhysRevD.100.103520,PhysRevD.94.123511} and a family of unified dark matter models (e.g., \cite{Camera:2019vbp} and references therein). Apart from the cosmological models, local gravitational potential \citep{PhysRevLett.110.241305}, specifically a local void \citep{Keenan13,Whitbourn14} can also partially relieve $H_0$ tension \citep{Hoscheit17,Shanks19}, yet there are studies utilizing SN data sets \citep{Kenworthy_2019,Lukovic19}, which show that the local structure does not significantly impact measurement of $H_0$.

Before we turn to revamp the standard $\Lambda$CDM model, it is necessary to get some insight from low redshift cosmological probes, whose variousness and observational accuracy can also provide us an integrated and precise understanding of the late universe. In this work, we perform a cosmological-model-independent reconstruction of $H(z)$, an inverse distance ladder analysis using the Type Ia Supernovae (SN Ia), Baryon  Acoustic Oscillations (BAO), and Gravitational Lens Time Delays (GLTD) data, which are able to impose a strong constraint on the shape of $H(z)$ and $H_0$ is simply obtained via extrapolation of $H(z)$ to present ($z=0$). 

GLTD provides a measurement of a combination of distances, when the lens mass model is assumed, the angular diameter distance to the lens can further be obtained. We include the GLTD data as it is an independent distance probe and is an excellent supplement to BAO and SN Ia, even though its current uncertainties are not comparable to the latter, it has the advantage of measuring the absolute distances, unlike, the SNIa, which need marginalization of the nuisance parameter, i.e, standardized absolute luminosity. 

Our analyses are closely related to the recent work by \citep{Lemos:2018smw} (hereafter \citetalias{Lemos:2018smw}), as we adopt the same parametric form of $H(z)$ and update the BAO data, include the GLTD data into analyses. We find that the reconstructed $H(z)$ nearly reproduces the one of the $\Lambda$CDM model. Our inferred $H_0$ when combining all three probes is slightly higher than the primary results of \citetalias{Lemos:2018smw}, which is mostly due to the inclusion of GLTD data, which predicts a higher $H_0$ than the \emph{Planck} $\Lambda$CDM estimate. Compared to \citetalias{Lemos:2018smw} we also include different priors on the parameters and different Bayesian statistical indicators 
to assess which models are preferred and the degree of degeneracy of the parameters.
As a more important extension, we forecast the performance of future BAO data from the Dark Energy Spectroscopic Instrument (DESI) \citep{Levi:2013gra} and GLTD data from the Large Synoptic Survey Telescope (LSST) \citep{Ivezic:2008fe}. The forthcoming data from these two future surveys are expected to provide a much tighter constraint on the reconstructed $H_0$.

The paper is organized as follows: In Section \ref{sec:meth} we introduce the parameterization methods of $H(z)$. In Section \ref{sec:data}, we present the data used to reconstruction as well as the inference method. We show the final results using the current and future data in Section \ref{sec:res} and then follow the discussion and summary in Section \ref{sec:sum}.

\section{Model and Equations}
\label{sec:meth}

 Firstly, we parameterize $H(z)$ in the following two ways: 
\begin{equation}
\label{eps}
\left(\frac{H(z)}{H_{0,\rm{fid}}}\right)^{2} = A_1(1+z)^3+B_1+C_1z+D_1(1+z)^\epsilon,
\end{equation}

\begin{equation}
\label{log}
\left(\frac{H(z)}{H_{\rm{0,fid}}}\right)^{2} = A_2(1+z)^3+B_2+C_2z+D_2\ln(1+z),
\end{equation}
which are the same as in \citetalias{Lemos:2018smw}, and denote them as Epsilon model and Log model, respectively. While these models serve the purpose of being blind to the dark energy equation of state, they are clearly inadequate to account for the curvature freedom. Moreover, ignoring the curvature would induce error that grows rapidly with redshift in reconstructing the dark energy equation of state \citep{Clarkson_2007}. To accommodate for this we also implement two additional models:
\begin{equation}
\label{poly}
\left(\frac{H(z)}{H_{0,\rm{fid}}}\right)^{2} = A_3(1+z)^3+B_3(1+z)^2+C_3+D_3\ln(1+z),
\end{equation}
\begin{equation}
\label{lcdm}
\left(\frac{H(z)}{H_{\rm{0}}}\right)^{2} = A_4(1+z)^3+B_4(1+z)^2+D_4.
\end{equation}

\begin{figure*}[htb!]
\begin{center}
\includegraphics[width=0.5\linewidth, clip]{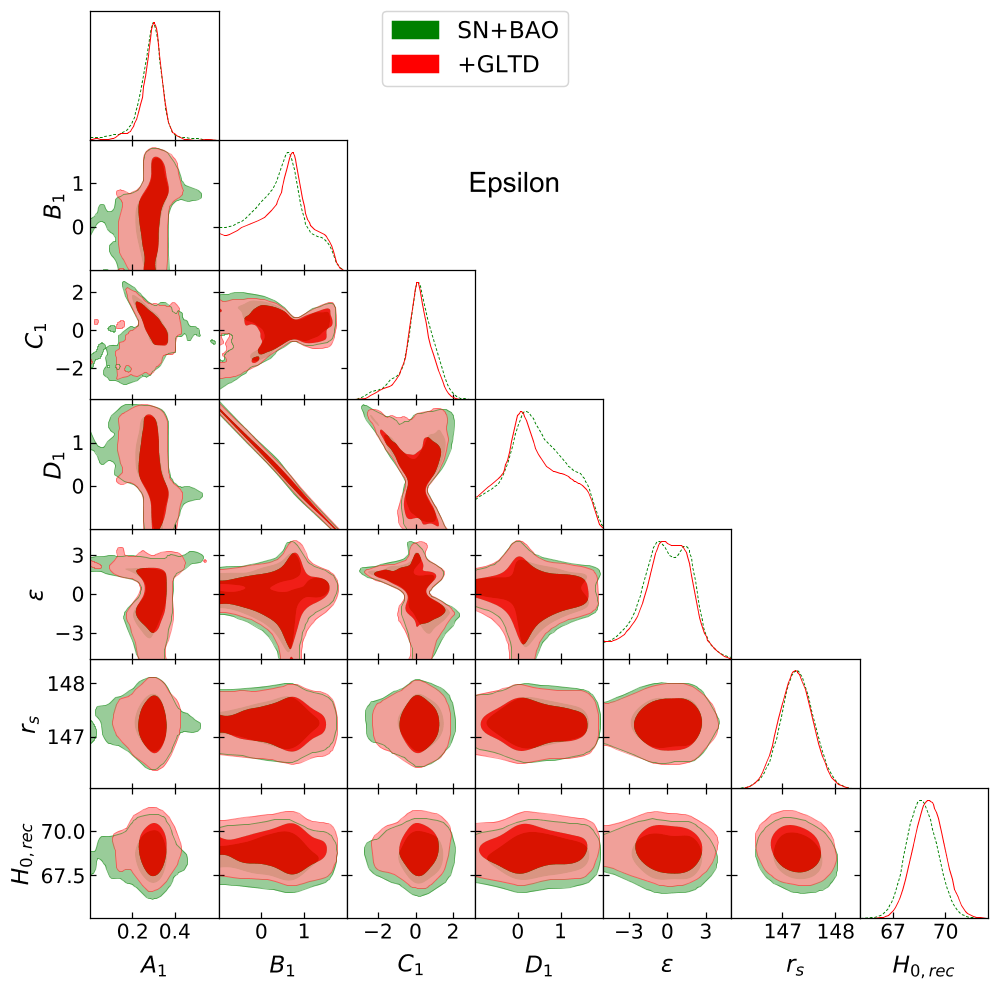}
\hspace{-0.07in}
\includegraphics[width=0.5\linewidth, clip]{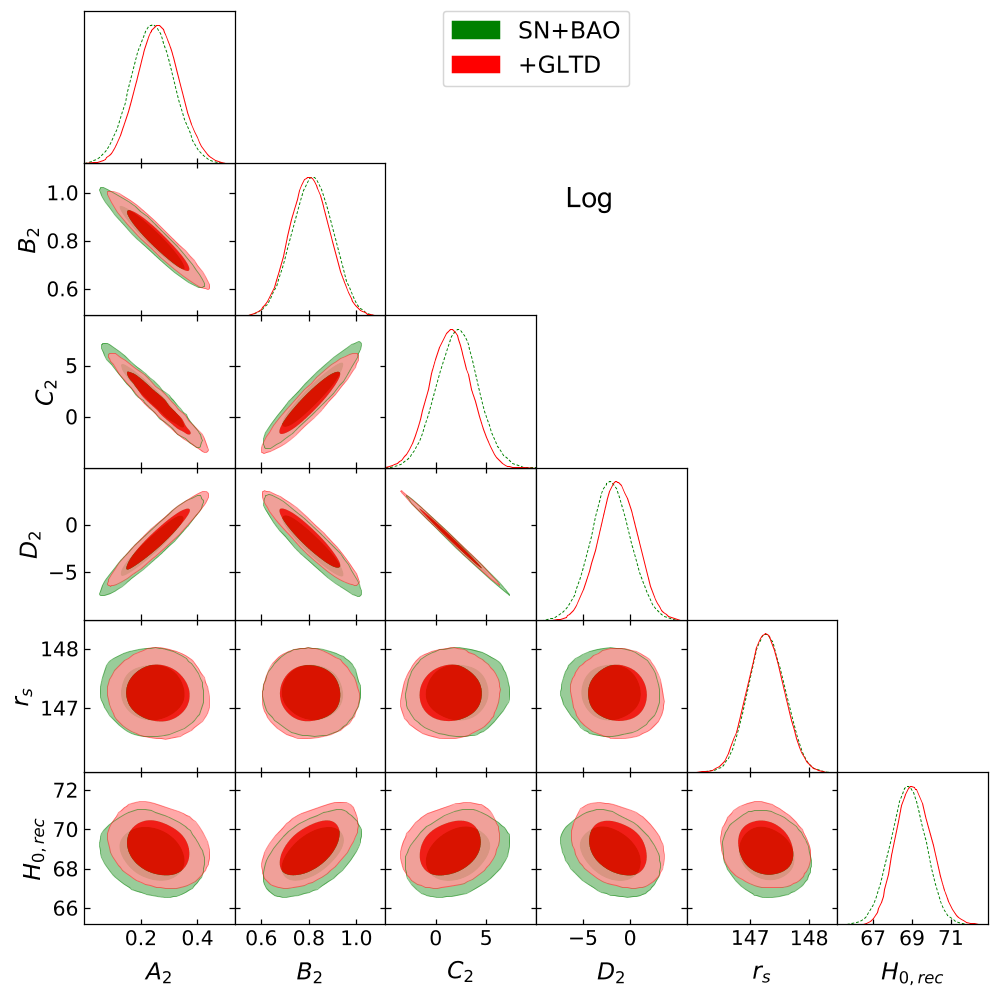}
\caption{{\it Left panel}: parameter constraints for the Epsilon model at 68\% and 95\% C.L. confidence level. {\it Right panel}: Same as left, but for the Log model. In both the panels we also show  $H_{\rm{0,rec}}$, which is a derived quantity.}

\label{epscon}
\end{center}
\end{figure*}

They are denoted as Log2 model and $\Omega_{\rm k}\Lambda$CDM model, respectively. We substitute the term $\propto z$ with a $\propto(1+z)^2$ term for theoretical and practical reasons: $i)$ the latter has cosmological implication for the curvature of the universe, $ii)$ as shown in right panel of \Cref{epscon}, the parameters $C_2$ and $D_2$ are strongly correlated, which is primarily due to $\ln(1+z)\approx z$ at small redshifts. We also implement the $\Omega_{\rm k}\Lambda$CDM model, which we write in a similar parametric form as the other models yet implementing restrictions on its parameters: $i)$ $H_0$ is a free parameter, which is a different implementation from other models where $H_{0,{\rm rec}}$ is a derived quantity, $ii)$ $A_4+B_4+D_4=1$, which is in fact the consistency relation when rewritten in terms of standard density parameters ($\Omega_{\rm m}+\Omega_{\rm k}+\Omega_{\Lambda}=1$). We adopt a fiducial Hubble constant value of $H_{0, {\rm {fid}}} = 67.0 \kmsMpc$. The reconstructed $H_0$, denoted as $H_{0,\rm{rec}}$, for each model is deduced at $z=0$ after extrapolation. The choice of $H_{0,\rm{fid}}$ does not alter $H_{0,\rm{rec}}$\footnote{We verify that a different assumption of $H_{0,\rm{fid}}$ hardly varies the inferred $H_{0,\rm{rec}}$ if we replace $H_{0,\rm{fid}} = 67.0 \kmsMpc$ with a different value, such as $H_{0,\rm{fid}} = 73.0 \kmsMpc$.}. 

\begin{table*}[ht]
\centering
\caption{Summary of BAO data used.}
\label{databao}
\renewcommand\arraystretch{1.05}
 \setlength{\tabcolsep}{2mm}{
\begin{tabular}{c|cccc}
\hline 
\hline  
Data set     &     $z_{\rm {eff}}$   &   Measurements   &   constraint & unit \\
\hline 
6dFGS     & 0.106    & $r_{\rm{s}}/D_V (z_{\rm {eff}})$  & $0.336 \pm 0.015$  &  $-$   \\
\hline 
\multirow{6}{*}{BOSS DR12 }&\multirow{2}{*}{0.38} &   $D_{\rm{M}} (z_{\rm {eff}})r_{\rm {s,fid}}/r_{\rm{s}}$ & $ 1512 \pm 25$&$ \rm{Mpc}$\\
                   &         &  $H(z_{\rm {eff}})r_{\rm{s}}/r_{\rm {s,fid}} $  &      $ 81.2 \pm 2.4 $&$\rm{km /s /Mpc}$\\
                   &  \multirow{2}{*}{0.51}  & $ D_{\rm{M}} (z_{\rm {eff}})r_{\rm {s,fid}}/r_{\rm{s}}$ & $ 1975 \pm 30 $&$\rm{Mpc}$ \\
                   &           & $H(z_{\rm {eff}})r_{\rm{s}}/r_{\rm{s,fid}}   $&   $90.9 \pm 2.3 $&$ \rm{km /s /Mpc}$\\
                    &   \multirow{2}{*}{0.61}  & $ D_{\rm{M}} (z_{\rm {eff}})r_{\rm {s,fid}}/r_{\rm{s}}$& $2307 \pm 37$&$ \rm{Mpc}$\\
                   &           &$ H(z_{\rm {eff}})r_{\rm{s}}/r_{\rm{s,fid}}   $&  $ 99.0 \pm 2.5 $&$\rm{km /s /Mpc}$\\
  \hline  
eBOSS DR14 QSO&1.52&$D_V r_{s,fid}/r_{\rm s}$ &$3843 \pm 147$&$ \rm{Mpc}$\\
\hline
\multirow{2}{*}{eBOSS DR14 LRG}& \multirow{2}{*}{0.72} &$D_{\rm{A}}(z_{\rm {eff}})r_{\rm{s,fid}}/r_{\rm{s}}$ & $1466.5\pm 136.6$&$ \rm{Mpc}$\\
                             &         & $H(z_{\rm {eff}})r_{\rm{s}}/r_{\rm{s,fid}}  $ & $105.8 \pm 16 $&$ \rm{km /s /Mpc}$\\
                             \hline  
\multirow{2}{*}{BOSS DR14 Ly$\alpha$}    & \multirow{2}{*}{2.34} & $D_{\rm{M}} (z_{\rm {eff}}) / r_{\rm{s}}$  &  $ 37.41 \pm 1.86$ &$-$ \\
                             &         & $ c/(H(z_{\rm {eff}})r_{\rm{s}})$&$8.86 \pm 0.29$&$-$\\
                             \hline  
\multirow{2}{*}{BOSS DR14 QSOLy $\alpha$} & \multirow{2}{*}{2.35}& $D_{\rm{M}} (z_{\rm {eff}})/r_{\rm{s}} $& $36.3 \pm 1.8$&$-$\\
                                   &       & $ c/(H(z_{\rm {eff}})r_{\rm{s}})$& $9.20 \pm 0.36$  & $-$\\
\hline
\end{tabular}
}
\end{table*}

\begin{table*}[ht]
\centering
\caption{Summary of GLTD data. Units of distances are all \rm{Mpc}.}
\label{tdl_data}
\renewcommand\arraystretch{1.05}
 \setlength{\tabcolsep}{2mm}{
 \renewcommand\arraystretch{1.05}
\begin{tabular}{c|cccc|ccc}
\hline 
\hline  
lens name     &     $z_d$   &  $z_s$   &   $D_{\rm{\Delta t}} (\rm{Mpc})$ &$D_{\rm A} (\rm{Mpc})$ & $\lambda$ & $\nu $ &  $ \sigma $ \\
\hline 
B1608+656&0.6304&1.394&$5156^{+296}_{-236}$&$ - $  &  4000   & 7.053   & 0.2282    \\
RXJ1131-1231&0.295&0.654&$2096^{+98}_{-83}$&$ - $  &  1388.8   &  6.4682  & 0.20560  \\
SDSS J1206+4332&0.7545&1.789&$5769^{+589}_{-471}$&$1805^{+555}_{-398}$&$-$&$-$&$-$\\
HE 0435-1223&0.4546&1.693&$2707^{+183}_{-168}$&$-$   &  653.9   &  7.5793  & 0.10312  \\
\hline 
\end{tabular}
}
\end{table*}

In both Log2 and $\Omega_{\rm k}\Lambda$CDM model, having the explicit $(1+z)^2$ term, which has the interpretation of cosmic curvature, the transverse comoving distance $D_{\rm{M}}$ becomes
\begin{equation}
\label{dm}
D_{\rm{M}}(z)=\begin{cases}
\frac{D_H}{\sqrt{\Omega_{\rm k}}}\sinh \left(\frac{\sqrt{\Omega_{\rm k}}D_C(z)}{D_H}\right), &\Omega_{\rm k}>0
\cr D_C(z), &\Omega_{\rm k}=0, 
\cr \frac{D_H}{\sqrt{-\Omega_{\rm k}}}\sin\left(\frac{\sqrt{-\Omega_{\rm k}}D_C(z)}{D_H} \right), &\Omega_{\rm k}<0 \end{cases}
\end{equation}
where the comoving distance $D_{\rm{C}}=c \int_0^z \frac{dz'}{H(z')}$ and $D_H=c/H_0$, $c$ is the speed of light. Thus, the luminosity distance $D_{\rm l}$ and angular diameter distance $D_{\rm{A}}$ are
\begin{equation}
D_L(z) = D_{\rm{M}}(z)(1+z),\, D_{\rm{A}} (z)= D_{\rm{M}}(z)/(1+z).
\end{equation}
BAO measurements often involve an effective volume averaged distance, denoted as $D_V$, and defined as:
\begin{equation}
D_V(z) = \left[ D_{\rm{M}}^{2}(z)\frac{cz}{H(z)}  \right].
\end{equation}

Based on \Cref{dm}, the comoving sound horizon $r_{\rm{s}}(z)$ at drag epoch is obtained by substituting the light speed $c$ with the sound velocity $c_s$ and changing the limit of integral from the early times ($z\rightarrow \infty$) to the drag epoch, $z_s$, which then reads:
\begin{equation}
\label{rd}
r_{\rm{s}}(z_s) =c_s \int_{z_{s}}^{\infty} \frac{dz'}{H(z')}\, ,
\end{equation}
where $c_s$ is a function of the ratio of baryon to photon energy density ($\rho_{\rm{b}}/\rho_{\rm {\gamma}}$), $c_s=1/\sqrt{3(1+3\rho_b/(4\rho_{\gamma}))}$. Our purpose here is to reconstruct $H(z)$ in a model-independent way, having minimum involvement with the physics of the early universe. Therefore, here we use a $r_{\rm s}$ prior from the \emph{Planck} \citep{Ade:2015xua}, which implies we assume the universe before $z_s$ is the same as depicted by the $\Lambda$CDM model. Also, it has been shown that the dark energy and curvature degree of freedom do not modify the expectation of $r_{\rm s}(z_s)$ \citep{Verde16, Verde17}. The $H(z)$ parameterizations in \Cref{eps,log,poly} are valid only in the late universe.

In a strong lens system, light from a background object is bent, maybe by an intervening mass (lens), and multiple images are generated. The lens systems usually show complicate morphologies and this implies that  the light rays go through different optical paths in the gravitational potential. In turn, this can be measured if the source has a variation in flux by relying on the difference in the arrival time, i.e., the time delay, of images. By measuring which, we finally obtain a combination of distance information of the lens system denoted as $D_{\rm{\Delta t}}$ \citep{1991ApJ...378L...5N,treu2016time}
\begin{equation}
D_{\rm{\Delta t}}=(1+z_{\rm l})\frac{D_{\rm l} D_{\rm s}}{D_{\rm ls}},
\end{equation}
where $z_{\rm l}$ is the redshift of the lens, $D_{\rm l}$ and $D_{\rm s}$ are the angular diameter distance from us to the lens and source, respectively. $D_{\rm{\Delta t}}$ has the dimension of distance and consequently is inversely proportioned to $H_0$. Moreover, with a proper assumption of lens mass density profile, one can extract $D_{\rm l}$ by combining it with the lens stellar velocity dispersion measurements and time-delay measurements \citep{paraficz2009gravitational,Jee:2014uxa}.

\begin{figure}[htb!]
\begin{center}
\includegraphics[width=1.\linewidth, clip]{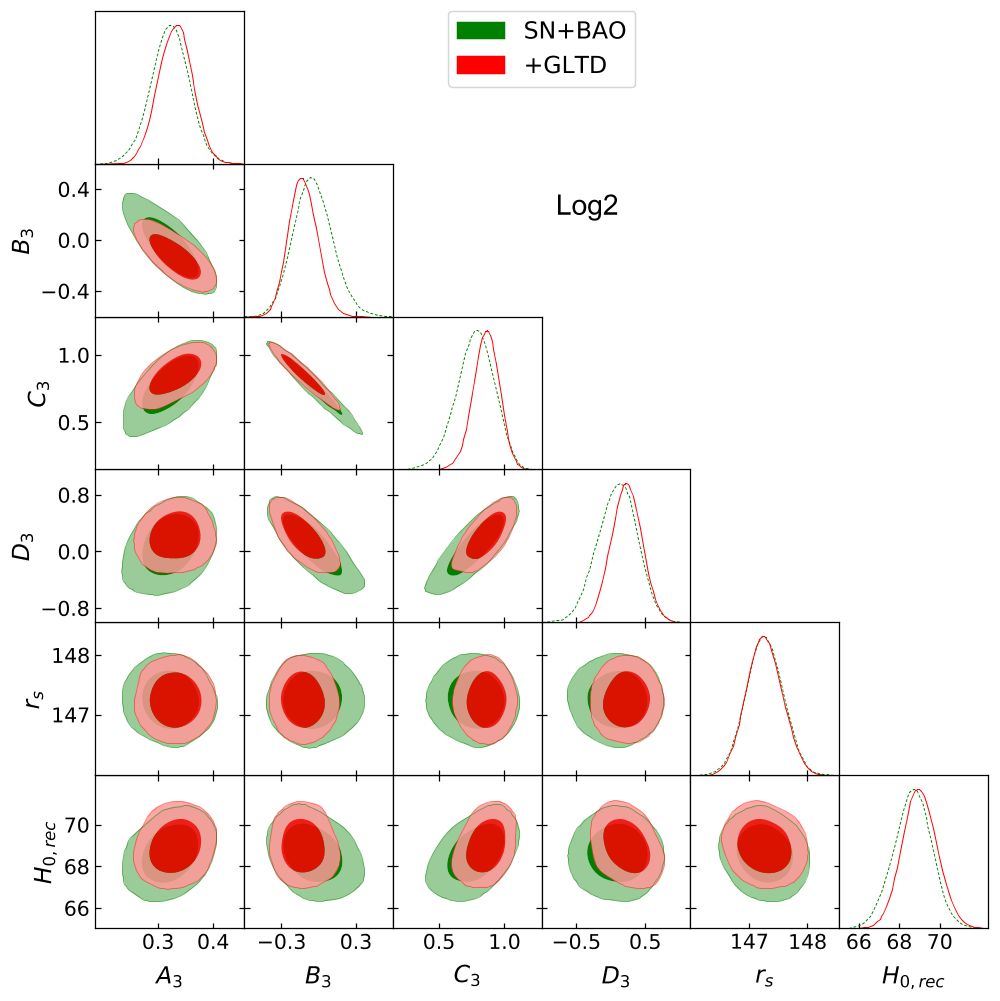}
\caption{Parameter constraints for the Log2 model at 68\% and 95\% C.L. limits. Here the parameter $B_3$ corresponds to curvature. We also show the reconstructed $H_{\rm{0,rec}}$.}
\label{log2con}
\end{center}
\end{figure}


\section{Data sets and Inference Method}
\label{sec:data}
Our work is mainly based on the following three probes: SN Ia, BAO, and GLTD. In this section, we summarize in detail the data used in the reconstruction of $H(z)$. Following which the inference method is also presented.
\subsection{Data sets}
\begin{itemize}[leftmargin=*]
\item SNIa from the new Pantheon sample \citep{Scolnic:2017caz}, contains a total of 1048 SN Ia spanning the redshift range from $0.01 < z < 2.3$. The Pantheon sample is a large combination of SN Ia from various surveys, including a subset of 279 SN Ia from the Pan-STARRS1 (PS1) Medium Deep Survey, SDSS, SNLS, various low-z, and HST samples. It has been widely used to constrain cosmology model and in particular, the nature of dark energy. For a given dark energy model, the Pantheon SNIa give consistent constraints on dark energy parameters with the results obtained using the joint light-curve analysis SNIa \citep{Betoule:2014frx} and also the latest Dark Energy Survey Supernova Program SNIa sample \citep{Abbott:2018wog}.
\item \Cref{databao} summarizes the latest BAO measurements used in our reconstruction. We use BAO measurements from 6dF Galaxy Survey (6dFGS) \citep{2011MNRAS.416.3017B} and BOSS DR12 in three redshift bands \citep{Alam:2016hwk}. The eBOSS DR14 also provides three high-redshift BAO measurements from quasar \citep{Zarrouk:2018vwy}, Lyman-$\alpha$ (Ly$\alpha$) absorption in the quasar spectrum \citep{Blomqvist:2019rah}, and Ly$\alpha$-quasar cross-correlation \citep{Agathe:2019vsu}. In addition, we as well use measurements on $D_{\rm{A}}(z_{\rm {eff}})/r_{\rm{s}}$ and $H(z_{\rm {eff}})r_{\rm{s}}$ from eBOSS DR14 luminosity red giants (LRG) analyses at $z_{\rm eff}=0.72$ \citep{Icaza-Lizaola:2019zgk}. We assume two measurements from eBOSS DR14 LRG are independent, as their covariance is unknown. Besides, due to the overlap of the CMASS sample, eBOSS DR14 LRG, and the last data point in BOSS DR12 would have a small correlation of $\sim 0.16$ \citep{Bautista18}. Both will lead to a very little influence on the reconstruction results and can hardly alter the constraint on $H_0$.

\item We use four GLTD distance measurements as summarized in \Cref{tdl_data}. The posterior likelihoods of the distance measures for GLTDs B1608+656 \citep{2010ApJ...711..201S}, RXJ1131-1231 \citep{Suyu:2013kha}, HE 0435-1223 \citep{Wong_2016} and SDSS J1206+4332 \citep{Birrer:2018vtm} are publicly available. The first three have robust measurements of the time-delay distance, given as skewed log-normal distribution $P(D_{\rm{\Delta t}}|\bm\theta)$:
\begin{equation}
\begin{aligned}
\label{sld}
P(D_{\rm{\Delta t}}|\bm \theta) &\approx\ \frac{1}{\sqrt{2\pi}\rm{(x-\lambda_D)\sigma_D}} \times\\
&\rm{exp}\left[-\frac{(\log(x-\lambda_D)-\nu_D)^2}{2\sigma^2_D}\right],
\end{aligned}
\end{equation}
where $\bm \theta=\{A_1,B_1,C_1,D_1, \epsilon ,r_{\rm s}\}$ for Epsilon model and $\bm \theta=\{A_2,B_2,C_2,D_2,r_{\rm s}\}$ for Log model etc., $\rm{x}$ is the model prediction of $D_{\rm{\Delta t}} \, ({\rm 1 \, Mpc})^{-1}$. $\rm{\lambda_D}$, $\rm{\sigma_D}$ and $\rm{\nu_D}$, which vary for different lenses are summarized in \Cref{tdl_data}. For J1206, the time-delay distance and angular diameter distance of the lens $D_{\rm l}$ are both provided, however, as samples of distributions available from the H0LiCOW website\footnote{\href{https://shsuyu.github.io/H0LiCOW/site/}{https://shsuyu.github.io/H0LiCOW/site/}}, for which a kernel density estimator based likelihood is implemented \cite{Birrer_2019} \footnote{In the flat $\Lambda$CDM model, our best fitting value of $H_0$ using J1206 alone is $69.94\pm5.58$ - $67.86\pm6.1$ depending on the flat prior of $\Omega_m$ and $H_0$, which is consistent with \cite{Birrer_2019}.}.

\item We impose a Gaussian prior on $r_{\rm s}$ according to the \emph{Planck} 2015 TT,TE,EE$+$lowP likelihood combinations \citep{Ade:2015xua},
\begin{equation}
r_{\rm s} = 147.27\pm0.31 \, \rm {Mpc}.
\end{equation}
We do not use the \emph{WMAP9} and the latest \emph{Planck} 18 prior because their $r_{\rm{s}}$ are consistent with \emph{Planck} 15 and the reconstruction results should not change as also manifested in \citetalias{Lemos:2018smw}. 
\end{itemize}

\subsection{Inference method}
The best-fitting value of the reconstruction parameters is obtained by minimizing the $\chi^2$ function using the Cosmological MonteCarlo (CosmoMC)\footnote{\href{https://github.com/cmbant/CosmoMC}{https://github.com/cmbant/CosmoMC}} \citep{Lewis_2002} and analyzed mainly using the GetDist package\footnote{\href{https://github.com/cmbant/getdist/releases/tag/1.0.0}{https://github.com/cmbant/getdist/releases/tag/1.0.0}. We also acknowledge the use of ChainConsumer package \citep{Hinton16}, available at \href{https://github.com/Samreay/ChainConsumer/tree/Final-Paper}{https://github.com/Samreay/ChainConsumer/tree/Final-Paper}.} \citep{Lewis19}. Given a Gaussian posterior likelihood function (PLF), the general form of $\chi^2$ is
\begin{equation}
\chi^2=-2\ln(\rm{PLF})=\bm{\delta}^{\dagger}\bf{C}^{-1}\bm{\delta},
\end{equation}
where $\bf{C}$ is the covariance matrix of the data, and $\bm{\delta}$ is the difference between the data and the theoretical predictions. The second expression is valid only when the PLF are Gaussian or approximately Gaussian. In case where PLF is skewed or non-Gaussian, such as the GLTD data mentioned in the Section \ref{sec:data}, then we use the first expression. The CosmoMC package has already included the likelihood source file for Pantheon and all BAO measurements except eBOSS DR14 LRG, and we use them directly. For the B1608, J1131, HE0435 GLTD data, we use the PLF described by \Cref{sld}. For the J1206, we first piece-wise divide the chain samples of $D_{\rm{\Delta t}}$ and $D_d$ into small bins. Then we calculate the discrete PLF in every 2-dimension bin, the following procedure being the same as for the other data.

We use the Deviance Information Criterion (DIC) to estimate the performance of the four models. DIC combines heritage both from Akaike Information Criterion and Bayesian Information Criterion and applies to parameter degeneracy \citep{Liddle:2007fy, spiegelhalter2002bayesian}. For a likelihood function $\mathscr{L}$, DIC is defined by
\begin{equation}
    \rm{DIC} = \overline{D(\theta)}+p_{\rm{D}}
\end{equation}
where $\rm{D(\theta)=-2\ln \mathscr{L}+C}$ and $\rm{p_D=\overline{D(\theta)}-D(\overline{\theta})}$. $\rm{C}$ is a constant that only depends on data. In this form, definition of DIC has a clear Bayesian interpretation that it deals with average of $\ln \mathscr{L}$ rather than the maximum values. Again, $\rm{p_D}$ also has its indication that it approximately equals to the effective number of parameters constrained by the data. If $\rm{p_D}$ is less than the number of free parameters of a model ($\rm{N_p}$), then it suggests that these parameters are highly degenerate. In \Cref{rrr}, we also list $\rm{p_D}$ for each of the models.

\begin{table}[htb!]
\centering
\caption{Summary of the priors imposed on free parameters for the four models.}
\label{prior}
\renewcommand\arraystretch{0.9}
 \setlength{\tabcolsep}{1.8mm}{
 \renewcommand\arraystretch{0.9}
\begin{tabular}{c|ccccc}
\hline 
\hline
model  & Epsilon &   Log   &    Log2  &  $\Omega_{\rm k}\Lambda$CDM \\
\hline
$A$    & $[0.0,2.0]$ &   $[0.0,2.0]$   &    $[0.1,0.6]$  &  $[0.1,1.0]$ \\ 
$B$    & $[-2.0,2.0]$ &   $[0.0,2.0]$   &    $[-0.6,0.6]$  &  $[-0.3,0.3]$ \\ 
$C$    & $[-5.0,8.0]$ &   $[-05.0,8.0]$   &    $[0.15,2.00]$  &   $0$ \\ 
$D$    & $[-2.0,2.0]$ &   $[-10.0,6.0]$   &    $[-1.0,5.0]$  &  $[0.5,1.2]$ \\ 
$\epsilon$    & $[-5.0,5.0]$ &   $-$   &    $-$  &  $-$ \\ 
$r_{\rm s}$ & $[130, 160]$ & $[130, 160]$& $[130, 160]$& $[130, 160]$\\
\hline
\end{tabular}
}
\end{table}

\section{Results and Discussion}
\label{sec:res}

\begin{table*}[ht]
\centering
\caption{Summary of the marginalized constraints on the reconstruction parameters and $r_{\rm s}$ with upper and lower uncertainties at 68\% confidence level. We impose flat prior on reconstruction parameters and Gaussian prior on $r_{\rm s}$. We also list ${\rm p_D}$, which is the effective number of parameters, $\Delta$DIC and $\Delta\chi^2$ w.r.t the $\Omega_{\rm k}\Lambda$CDM model. All derived quantities are indicated with $^{*}$. For the reference model we show the DIC and $\chi^2$, for which $\Delta\rm{DIC} = \Delta\chi^2 = 0$.}
\renewcommand\arraystretch{1}
\setlength{\tabcolsep}{1.8mm}{
\renewcommand\arraystretch{1}
\begin{tabular}{c|cc|cc|cc|cc}
\hline 
\hline  
model & \multicolumn{2}{c}{Epsilon} & \multicolumn{2}{c}{Log} & \multicolumn{2}{c}{Log2}&\multicolumn{2}{c}{$\Omega_{\rm k}\Lambda$CDM} \\
\hline 
Data set &  SN+BAO   &  +GLTD   &     SN+BAO   &  +GLTD  &   SN+BAO   &  +GLTD &   SN+BAO   &  +GLTD \\
\hline 
$A$ & $0.29_{-0.05}^{+0.06}$ & $0.31_{-0.06}^{+0.04}$ &  $0.24_{-0.07}^{+0.07}$ &$0.26_{-0.07}^{+0.07}$ & $0.32_{-0.03}^{+0.03}$ &$0.33_{-0.03}^{+0.03}$ & $0.30_{-0.03}^{+0.03}$ &$0.31_{-0.03}^{+0.03}$ \\
$B$ & $0.31_{-0.87}^{+0.72}$  &  $0.45_{-0.88}^{+0.75}$ &$0.81_{-0.09}^{+0.09}$  & $0.80_{-0.08}^{+0.08}$ & $-0.05_{-0.17}^{+0.14}$ &$-0.13_{-0.12}^{+0.11}$ & $0.01_{-0.10}^{+0.09}$ &$-0.05_{-0.08}^{+0.08}$ \\
$C$ & $0.01_{-0.61}^{+1.91}$&$0.00_{-0.65}^{+0.92}$  &  $2.11_{-2.08}^{+2.07}$ &  $1.40_{-2.03}^{+2.04}$ & $0.77_{-0.13}^{+0.15}$ &$0.86_{-0.09}^{+0.11}$ & $-$ &$-$ \\
$D$ & $0.46_{-0.79}^{+0.93}$ &  $0.30_{-0.93}^{+0.81}$ &$-2.11_{-2.09}^{+2.12}$   &  $-1.37_{-2.07}^{+2.07}$ & $0.09_{-0.27}^{+0.30}$ & $0.22_{-0.21}^{+0.23}$ & ${0.69_{-0.06}^{+0.07}}^{*}$ &${0.74_{-0.06}^{+0.06}}^{*}$ \\
$\epsilon$ & $-0.02_{-1.42}^{+2.20}$  & $0.11_{-1.46}^{+2.09}$  & $-$  &  $-$ & $-$  &  $-$ & $-$  &  $-$ \\
$r_{\rm s}$ & $147.26_{-0.31}^{+0.31}$  & $147.24_{-0.31}^{+0.31}$  &$147.28_{-0.31}^{+0.31}$    &  $147.23_{-0.31}^{+0.31}$ & $147.26_{-0.32}^{+0.32}$    &  $147.25_{-0.30}^{+0.30}$    &   $147.25_{-0.32}^{+0.31}$    &  $147.24_{-0.30}^{+0.30}$  \\
$H_0$ & ${68.62_{-0.89}^{+0.89}}^{*}$  & ${69.01_{-0.85}^{+0.84}}^{*}$  & ${68.77_{-0.90}^{+0.90}}^{*}$    &  ${69.13_{-0.90}^{+0.90}}^{*}$ & ${68.64_{-0.89}^{+0.96}}^{*}$    &  ${69.04_{-0.86}^{+0.86}}^{*}$ &  $  {68.59_{-0.95}^{+0.93}}$    &  ${69.11_{-0.98}^{+0.84}}$ \\
\hline
${\rm p_D}$      &5.16 &4.52 &4.90 &4.88 &5.11 &4.91 &4.02 &3.87\\
$\Delta$DIC(DIC)    &$+0.82$ &	$+1.09$ &$+0.56$&$+1.91$ &	$+2.00$ &$+1.12$&$ 1047.80$ &$1053.85$\\
$\Delta\chi^2$($\chi^2$) &	$-1.47$ &$-0.22$ &	$-1.21$ & $-0.12$ &	$-0.20$ &$-0.97$ &$1039.77$ &	$1046.12$\\

\hline
\end{tabular}
\label{rrr}
}
\end{table*}

We assume flat priors on the free parameters, as summarized in \Cref{prior}. The constraint results are presented in \Cref{rrr} and graphically in \Cref{epscon,log2con,hz} and the mock results are shown in \Cref{mock}.

\subsection{Constraints from current data}

We first use the most recent BAO and Pantheon SN Ia, the constrained results for the Epsilon and Log model are consistent with those reported in \citetalias{Lemos:2018smw}, with a mild improvement in the accuracy of the parameters due to the newer BAO data. When including the GLTD data, we find no tightening of the constraints, with a mild shift in the marginalized PLF of parameters globally. 

The Log model shows highly correlated, however much simpler, Gaussian constraints  than the Epsilon model which demonstrates a high degeneracy between the parameters. This degeneracy in the Epsilon model is driven by the parameter $\epsilon$, with a double peak in the marginalized posteriors. In comparison to the results of \citetalias{Lemos:2018smw}, we notice that the double peak behavior of $\epsilon$ is diminished when the prior on $B_1,\,D_1$ parameters are extended to negative ranges and completely vanishes when the GLTD data is included, as can be seen in \textit{Left panel} of \Cref{epscon}.

\begin{figure*}[t]
\begin{center}
\includegraphics[width=.95\textwidth,height=10cm]{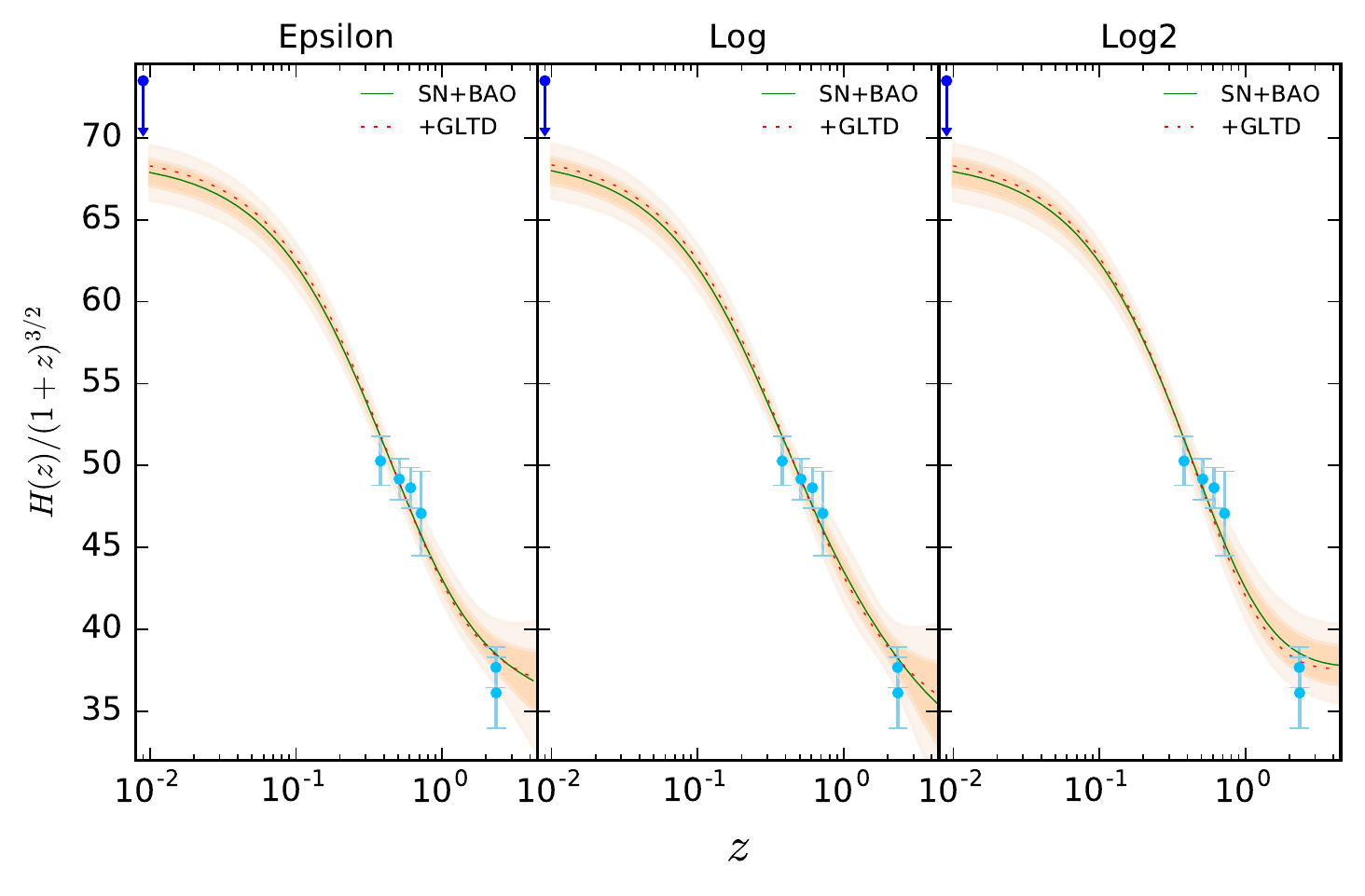}
\caption{$H(z)$ reconstruction results. The shaded region is the 1$\sigma$ and 2$\sigma$ error range of the joint constraint from Pantheon+BAO+GLTD. The light blue points are the BAO estimates of $H(z)$ with its 1$\sigma$ error. The blue point is the local $H_0$ measurements and its lower 2$\sigma$ limit from the distance ladder \citep{Riess_2018}.}

\label{hz}
\end{center}
\end{figure*}

The BAO data when combined with the large number of SNIa samples, places tight constraints on the shape of $H(z)$. However, the correlations of the posteriors are compelling, which indicates substantial redundancy of these parameters. To this end, we also estimate the effective number of parameters ($\rm{p_{D}}$) constrained, which for the Epsilon model is always less by $\sim 1$, than the number of free parameters in the likelihood analysis. As shown in \Cref{rrr}, for the other three models implemented here, the effective number of parameters is almost equivalent to the number of free parameters. This in-turn is one of motivations to utilize the Log model to perform the mock analysis, elaborated later. 

\Cref{hz} describes the evolution of reconstructed $H(z)$, with the $z-$axis in log scale in the limits $z\lesssim 4$. We notice that the Epsilon model in fact extends to negative values of $H(z)$, when extrapolated to larger redshifts. When including GLTD, a rise of $H(z)$ in the low redshift range appears for all models, which results in a slightly higher $H_{\rm{{0,rec}}}$ of the order $\Delta H_{\rm{0,rec}} \sim 0.5$. This is in accordance with the fact that $r_{\rm s}$ prior with the BAO data supersedes the precision with which the GLTD data constrain the present expansion rate.

\begin{figure}[t]
\begin{center}
\includegraphics[width=1.\linewidth, clip]{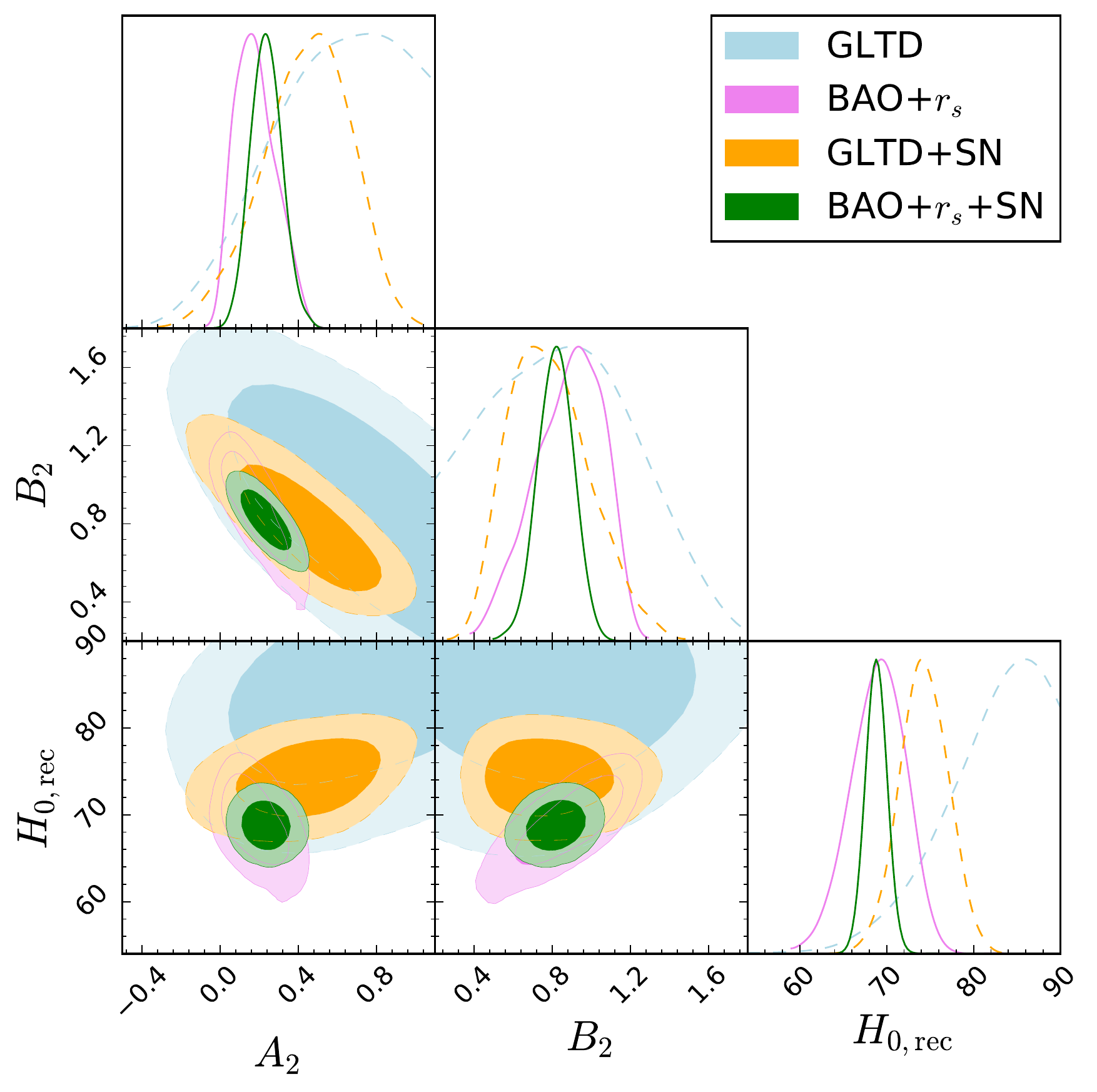}
\caption{Comparison of constraint results of different data sets for the Log model. All parameters of Log model are free. Here we only report parameters of most interest, $A_2$,\,$B_2$ and $H_{0,rec}$ (see the discussion in the text).}
\label{data_com}
\end{center}
\end{figure}

Both GLTD and BAO$+r_{\rm s}$ can independently determine $H_0$ while their inference shows mild discrepancy in the flat $\Lambda$CDM model \citep{Aghanim:2018eyx,Wong19}. We plot the constraints from GLTD and BAO$+r_{\rm s}$, as well as their respective combination with SN, i.e., GLTD$+$SN and BAO$+r_{\rm s}+$SN for the Log model in \Cref{data_com}. As expected, the constraints from GLTD on the model parameters are far less stringent. However, it is sufficient to constrain three parameters of interest: $A_2$, $B_2$ and $H_{\rm{{0,rec}}}$. The constraints from GLTD are consistent with other data sets well within the 1$\sigma$ region, for the first two parameters. As for the inferred $H_{\rm{{0,rec}}}$, we find a mild tension between GLTD and BAO$+r_{\rm s}$. When combined with SN, both, i.e., GLTD$+$SN and BAO$+r_{\rm s}+$SN data sets prefer lower $H_{\rm{{0,rec}}}$ values, while the tension remains since their error bars shrink as well. As shown in \Cref{data_com}, when contrasting the constraints form BAO$+r_{\rm s}$ (pink) against SN$+$GLTD (orange), it is noticeable that the correlation between parameter $A_2$, which scales as the matter density and $H_{\rm{{0,rec}}}$, is negative (i.e, $A_2\xrightarrow{}0$, for higher values of $H_{\rm{{0,rec}}}$) for the former and positive for the latter data set. This in fact results in a lower value of $H_{\rm{{0,rec}}}$ in the joint analysis and demonstrates why a low-redshift modification, as in the case of a parametric Log model cannot resolve the $H_0$-tension. Similar behavior was also earlier noted in \cite{Bernal16} (see Table 4. therein), using spline based reconstructions, where the SN data along with an $r_{\rm s}$ prior disfavored higher values of $H_0$, also validating the adequate utility of parametric reconstructions employed here. 

Preference for a higher (w.r.t CMB) value of Hubble constant from GLTD is clearly in line with other reports (e.g, Figure 2 and Table 5 in \citet{Wong19}), also in cases where the Hubble constant is determined via calibrated SN using absolute distances from GLTD \citep{Jee:2019hah}. However, due to its larger uncertainty, at present, it hardly plays a significant role in determining $H_{\rm{{0,rec}}}$, in a joint analysis with BAO data. The most recent GLTD data contain 6 gravitationally lensed quasars with updated measurements on both $D_{\rm{\Delta t}}$ and $D_{\rm{d}}$ \citep{Wong19}, 
for which, the constraints could become even tighter and consequently the tensions could be more even more significant\footnote{We were unable to implement the 6 GLTD data set from \cite{Wong19} here, as they are not yet made publicly available.}. 

Next, we consider the models Log2 and $\Omega_{\rm k}\Lambda$CDM (also the reference model), which have a curvature term in their parametric expressions. \Cref{log2con} shows the constraint contours for Log2 model, which are quite similar but with a reduced degeneracy in comparison to the Log model. This is in effect due to the replacement of the linear term with the quadrature term, which now plays the role of curvature. When including the GLTD data, we find a negative curvature parameter $B_3$, and a larger value for constant parameter $C_3$, to be compared with the $B_2$ parameter of the Log model. The effect on the value of $H_{\rm{{0,rec}}}$, is similar to that in the Log2 and the Epsilon models. The shape of $H(z)$ for the Log and Log2 model show a major difference at high redshifts, where the Log model falls faster with its error bars tending to diverge. While it is not visible when plotting with the $-$axis in logarithmic, we find that the Log model is, in fact, better driven by the data, which is not the case for the Log2 model whose $H(z)$ evolves more gradually at both extremes of redshift range. This data driven behavior also affirms the aforementioned motivation based on effective number of constrained parameters, to utilize the Log model to perform the mock analyses.

All the numerical results are summarized in Table \ref{rrr}, along with three statistical quantities for the model selection, which are the effective number of model parameters $\rm{p_D}$, DIC, and $\chi^2$ at best-fitting value. While $\rm{p_D}$ is a part of DIC estimate, we list it separately as it estimates the number of parameters of the model that are adequately constrained by the data. For instance, the Epsilon model has the most complicated degeneracy among the parameters of the model. Thus it is expected (and indeed) to have a smaller $\rm{p_D}$ than the number of free parameters (i.e, 6). We further find that for every model, $\rm{p_D}$ always becomes smaller after GLTD is included, which is mostly due to the fact that GLTD is in mild tension with SN and BAO, as shown in \Cref{data_com}. Including GLTD would actually increase the freedom, i.e., the degeneracy of free parameters allowed solely by SN or BAO. For the Log model alone we find almost no variation in $\rm{p_D(N_p)}\sim 4.9(5)$, with the inclusion of the GLTD data set, also being very close to the number of free parameters in the likelihood analysis.

The constraining ability of combined datasets on the four models is similar, having negligible difference in $H_{\rm{0,rec}}$ estimates. However, we notice that the Log model provides slightly conservative constraints on $H_{\rm {0,rec}}$, owing to a different behavior with the GLTD dataset. For the Epsilon and Log models, using GLTD alone we find $H_{\rm{0,rec}} = 80.9\pm6.7 \kmsMpc$ and  $H_{\rm{0,rec}} = 85.1\pm7.3 \kmsMpc$, respectively. Our constraint for the Epsilon model is more similar to the constraint from $w\neq-1$ extension of $\Lambda$CDM using the same dataset, recently reported in \cite{Taubenberger19}. The $\Omega_{\rm k}\Lambda$CDM model is the most optimal fit with the smallest DIC, essentially due to the smallest number of free parameters, having similar $\chi^2$ values to the other models. For instance, with the Log model, $H_{\rm{0,rec}}$ is obtained by extrapolating the reconstructed $H(z)$ to $z=0$, for which we find 68\% C.L. limits of,
\begin{align*}
H_{\rm{0,rec}}&=68.77^{+0.90}_{-0.90} \,\kmsMpc\,(\rm{SN+BAO}),\\
H_{\rm{0,rec}}&=69.13^{+0.90}_{-0.90} \,\kmsMpc\,(\rm{SN+BAO+GLTD}).
\end{align*}
\hypertarget{tension}{These values of $H_0$ in comparison to the \emph{Planck} $\Lambda$CDM\footnote{For the same \emph{Planck} likelihood combination utilized for $r_{\rm s}$ prior here, the corresponding 68 \% C.L. limit is $H_{\rm{0}}=67.27\pm 0.66 \kmsMpc$, for the $\Lambda$CDM model.} and local measurement\footnote{We assume the value of $H_0=73.45\pm{1.66} \kmsMpc$, from \cite{Riess_2018} (hereafter \citetalias{Riess_2018}).} are at $\sim 1.3\sigma$\footnote{{As is the usual practice in an inverse distance ladder comparison, we assume no correlation between our $H_{\rm 0,rec}$ and \emph{Planck} $H_0$, however, $r_{\rm s}$ prior is strongly ($+0.79$) correlated to the latter and our $r_{\rm s}$ posterior is mildly ($-0.14$) anti-correlated with the former while being equivalent to the prior. Implying $\sim -0.12$ anti-correlation between the two $H_0$ quantities and is expected to increase the deviation and might have a role to play with more precise future data, for instance, increasing to $-0.27$, in the forecast analysis presented in Section \ref{sec:res:fut}.}} using SN$+$BAO ($ 1.7\sigma$ using SN$+$BAO$+$GLTD) and $\sim 2.5\sigma$ ($2.3\sigma$), respectively. In the earlier analysis, \citetalias{Lemos:2018smw} quote a $1.0\sigma$ and $2.7\sigma$ for the same comparison with SN+BAO data.} However, when the GLTD data are included, our inferences for the respective tensions move in the direction of the results presented in \cite{Dutta19}, whose analyses include Cosmic Chronometers (CC) \citep{Jimenez02, Moresco15} and growth measurements from large scale structure observations.

Although the value of  $H_{\rm{0,rec}}$ is slightly raised by GLTD, they are too small to be consistent with the local measurement of $H_0$. While this situation would change if the GLTD becomes more accurate and precise, at the current stage, our reconstructed Hubble parameter still favors the \emph{Planck} estimate and is in agreement with other earlier analyses \citep{Aubourg_2015, Bernal16, Feeney17}. Incidentally, we also notice that the $H_0$ estimates in our analyses, driven by the combination of GLTD data and $r_{\rm s}$ prior are extremely consistent with those reported in \citep{Haridasu18, Gomez-Valent18, Mukherjee19}\footnote{See for example, other works driven by CC based $H_0$ estimations \citep{Lukovic16, Yu17, Lukovic18, Park19}, which at times do not account for the systematics within CC data.}, which were driven by CC datasets. These earlier results are also model-independent, being very different from the approach implemented here. The low-redshift model-independent (see e.g., \cite{Haridasu18}) constraint on the compound parameter $r_{\rm s} \times H_0/[100 \kmsMpc] \,(r_{\rm s}h)$ is consistent with the \emph{Planck} estimate, even with the inclusion of local $H_0$ \citep{Riess_2018}, within $1\sigma$. And also, in line with the earlier analysis performed in \cite{Carvalho:2015ica}, we replace the $r_{\rm s}$ prior with $r_{\rm s}h = 99.069 \pm 1.598 \, \rm{Mpc}$ prior, obtained from the same combination of \emph{Planck} likelihood. This clearly allows for a larger value of $H_0 = 73.86\pm 2.41 \, \kmsMpc$ and a corresponding $r_{\rm s} =136.0 \pm 4.5\, \rm{Mpc}$, consistent with \cite{Arendse19} as expected, and is accompanied by a change in the best-fitting $\chi^2$ value for GLTD data by $\sim 3$, while the same for SN and BAO data sets remain almost unchanged. In contrast to the $r_{\rm s}h$ prior, when $r_{\rm s}$ prior is implemented, as in the main analysis, the posterior estimate of $r_{\rm s}h = 101.76\pm1.32 \, \rm{Mpc}$, is driven towards larger values, and consistent with the \emph{Planck} $r_{\rm s}h$ prior, at $\sim1.3\sigma$, which is a mild reduction in the $H_0$ alone \hyperlink{tension}{$\sim1.7\sigma$} deviation mentioned earlier. 

Please note that the $r_{\rm s}$ prior alone might ensure that the early Universe evolution is fixed to $\Lambda$CDM, as any of the one parameter extension such as, $\Omega_{\rm k}\neq 0$ or $w\neq -1$, would have the same $r_{\rm s}$ (i.e, same early-time behavior, also validating our use of same prior for the Log2 and $\Omega_{\rm k}\Lambda$CDM models with curvature freedom), but with a different late-time $H_0$ \citep{Ade:2015xua,Verde17}, consequently a different $r_{\rm s}h$, w.r.t $\Lambda$CDM. This in fact indicates that the early-time behavior constrained from the CMB data while being invariant for such extensions, would imply that the deviations are mainly enhanced when the models are extrapolated to late-time expansion history. However, an $r_{\rm s}h$ prior from the $\Lambda$CDM fit to the CMB data, would necessarily imply a \textit{correlated} early and late time behavior, also allowing for a possibility to break the $r_{\rm s} - h$ degeneracy differently. As already mentioned, an agreement for the constraint on $r_{\rm s}h$ from low-redshift BAO and high-redshift CMB, alongside the conformity of higher (than CMB) $H_0$ values from local distance ladder (\citetalias{Riess_2018}, \cite{Riess:2019cxk}) and GLTD \citep{Wong19} data sets\footnote{Please see \cite{Verde:2019ivm} and \cite{Riess19_Nat} for a summary of other low-redshifts probes which imply similar $H_0$ estimates.}, taken at a face value (assuming no spurious systematics) would indicate a need for modification of early-time physics. {One might tentatively infer that, while an early universe modification as a solution for the $H_0$-tension is desirable, such a change should necessarily be accompanied with a conserved/invariant $r_{\rm s}h$ (w.r.t $\Lambda$CDM) estimate from CMB, placing an additional restraint on feasible modifications. To this end, the comparison of $r_{\rm s}$ and $r_{\rm s}h$ prior analyses helps to assess the extent of allowed variation in the CMB $r_{\rm s}h$ estimate, from the low-redshift BAO data (also aided by SN). A modification that requires a change in $r_{\rm s}h$, would also imply a change in angular scales at recombination, which are very well constrained by CMB and subsequently effect the BAO observables, through the assumed fiducial cosmology.} {In this context, the BAO + Big Bang Nucleosynthesis (BBN) $H_0$ estimate has been shown to be consistent with the CMB estimate \citep{Aubourg_2015, Addison17, Blomqvist:2019rah, Schoneberg19,Cuceu_2019}, also in \citetalias{Lemos:2018smw}, and hints for a modification requiring a change in the $r_{\rm s}h$ estimate from CMB, which when implemented through the fiducial cosmology in obtaining/rescaling BAO observables, can allow reconciliation with the local $H_0$ estimate (see also \cite{Camarena19}).}

\subsection{Constraints from future data}
\label{sec:res:fut}
While the analysis so far, with the up-to-date BAO and GLTD data reaffirms the inferences of \citetalias{Lemos:2018smw}, we now more importantly forecast the constraining ability of realistic future BAO and GLTD data sets on $H_0$, through the model-independent formalism. While several future surveys such as Euclid \citep{Amendola:2016saw} and the Square Kilometre Array \citep{Bacon:2018dui} can provide precise measurements on BAO \citep{Bengaly:2019oxx,Obuljen_2018}, here we focus on BAO from DESI. And GLTD from Large Synoptic Survey Telescope (LSST). 

DESI is a Stage IV ground-based experiment started in 2019\footnote{\href{https://www.desi.lbl.gov/}{https://www.desi.lbl.gov/}}. It aims at studying BAO and the growth of structure through measuring spectra from 4 target tracers, including luminous red galaxies up to $z\sim1.0$, bright [O II] emission line galaxies up to $z\sim1.7$, quasars and Ly-$\alpha$ forest absorption feature in their spectrum up to $z=3.5$. Following \citet{Aghamousa:2016zmz}, we use the forecasted BAO measurements, which are quoted as $D_A(z)/r_{\rm s}$ and $H(z)r_{\rm s}$, from DESI galaxy, quasar and bright galaxy survey and also assume a correlation coefficient of 0.4 between these two measurements at each redshift. 

\begin{figure}[htbp]

\begin{center}
\includegraphics[width=1.\linewidth, clip]{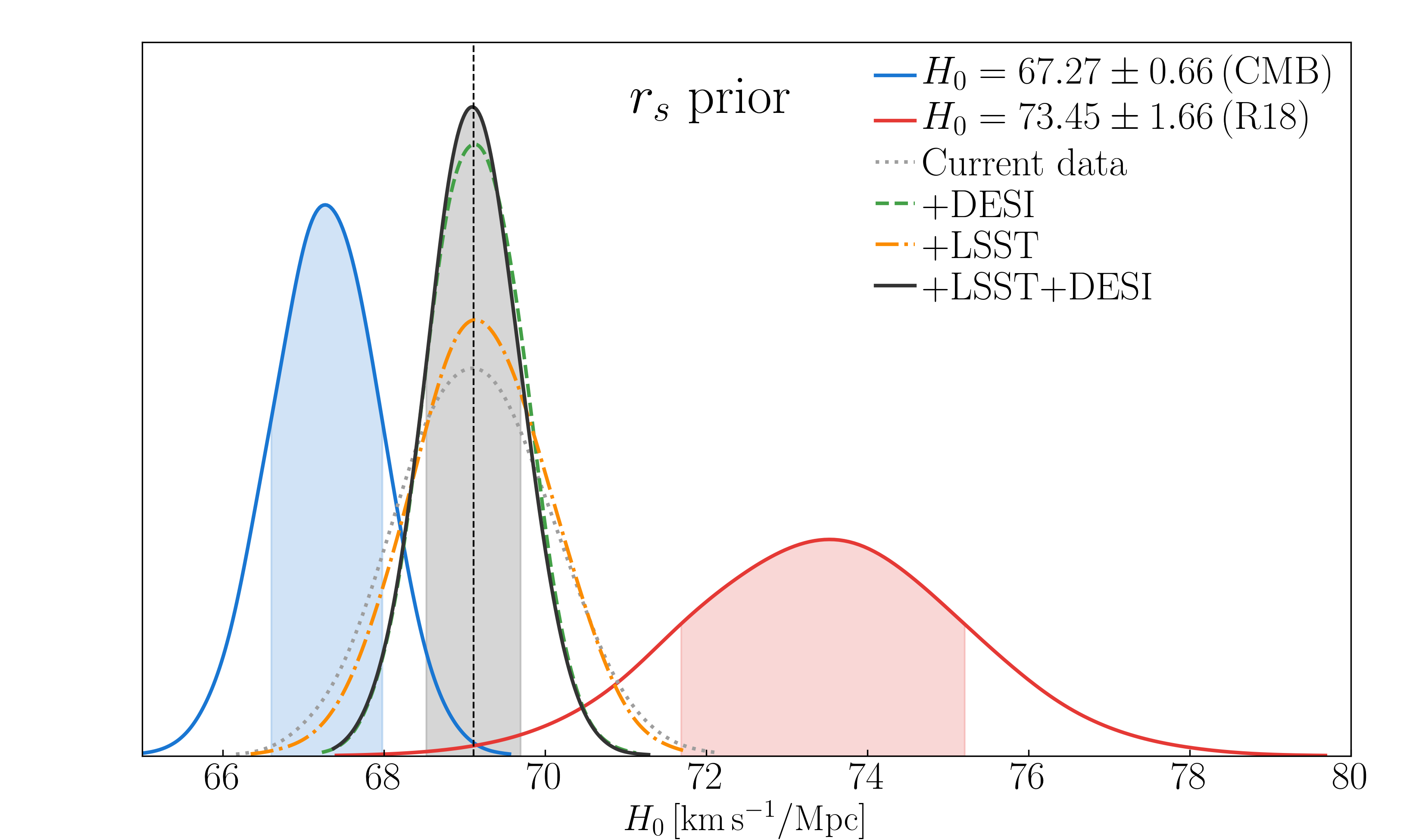}
\includegraphics[width=1.\linewidth, clip]{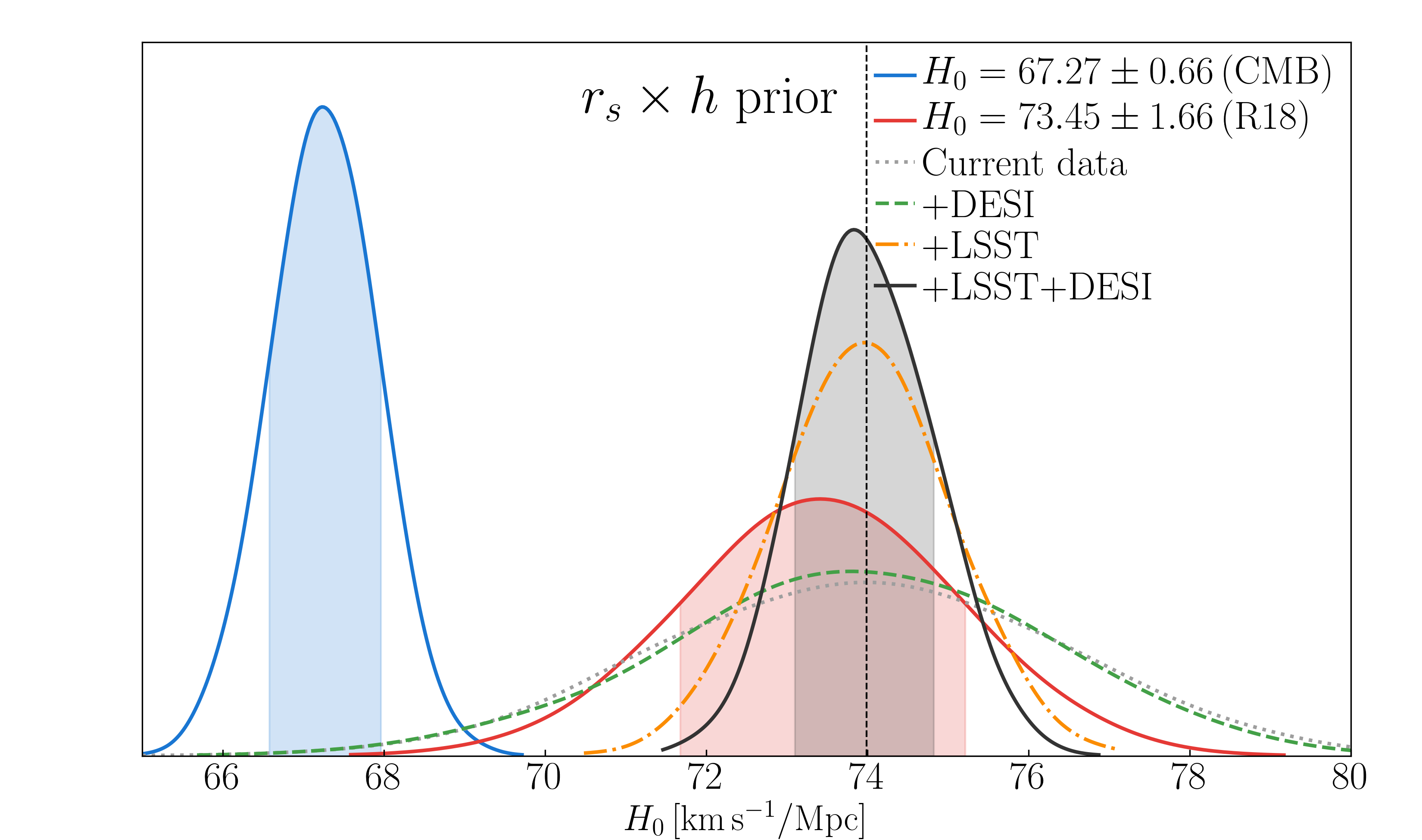}
\caption{Forecasts of marginalized $H_0$ using the future data, i.e., BAO from DESI and GLTD data from LSST. The upper panel uses the $r_{\rm s}$ prior and lower panel implements the $r_{\rm s}h$ prior, both taken from the same \emph{Planck} likelihood. We choose the fiducial model following the best-fitting of the joint constraint from Pantheon SNIa, BAO and GLTD (i.e., current data). The vertical dashed line represents the mean value from the posterior. It is important to stress the reversal in constraining ability of mock BAO and GLTD data sets, when changing from $r_{\rm s}$ to $r_{\rm s}h$ prior.}

\label{mock}
\end{center}
\end{figure}

LSST is an ambitious wide-deep-fast sky survey that plans for regular survey operations by 2022 \citep{Ivezic:2008fe}. \citet{Oguri_2010} made predictions of the numbers of time-variable sources that should be found by LSST and reported a very positive result that around 3000 of lensed quasars will have well-measured time delays. Based on the catalog of mock lenses in \citet{Oguri_2010}, \citet{Jee_2016} further forecasted the cosmographic constraints when including both $D_{\rm{\Delta t}}$ and $D_{\rm l}$ of lens systems. As the authors anticipated, there should be $\sim 55$ high-quality quadruple lens systems that have sufficiently good measurements of both distance information. However, this number may vary due to various limitations, for example, telescope observation strategy \citep{Liao:2019aty}. Furthermore, there should be a correlation between the measured $D_{\rm{\Delta t}}$ and $D_{\rm l}$ estimates or otherwise one of the distances should have much large uncertainty. Due to the lack of correlation information, in their paper, here we assume that only $D_{\rm{\Delta t}}$ is available. According to the current four GLTD data, the uncertainty on $D_{\rm{\Delta t}}$ varies within $\sim 5.8\%-7.0 \%$. Hence, a $5\%$ uncertainty level is achievable as long as we select the lens systems following the same criteria as \citet{Jee_2016}. The number of forecasted lens systems is conservatively reduced to 40.

We use the distribution of source and lens redshifts from \citep{Jee_2016} and randomly generate 40 lens systems. In principle, the 40 systems produced every time will have mildly different constraining ability depending on the redshift distribution of lenses and sources. We experimentally tested the fluctuation in the expected $\lesssim 1\sigma$ error by repeating MCMC analyses using different sets of mock GLTD data. We find the variation is much smaller than the uncertainty of the inferred $H_0$\footnote{ We run 20 separate MCMC analyses and find that the variation in the uncertainty of inferred $H_0$ relative to the corresponding mean is $\sim 2.5\%$, which should also contain the MCMC sampling noise.}. Thus, we use the one-run simulation results as a quantitative estimate of the constraining ability. 

The top panel of \Cref{mock} shows the 1D marginalized posterior of inferred $H_0$ when combining the current data with the future BAO and GLTD data for the Log model, with the $r_{\rm s}$ prior, where the relative heights are also indicative of the constraining ability of the data. For convenient comparison, we plot the current constraint in dotted gray. We do not analyze the other three models in detail, as they are not expected to exhibit much difference, which we verify and that the improvement in percentages will remain the same. However, testing the Epsilon model we find that it is less reliable to reproduce the model utilized to create the mock data set, due to stronger intrinsic degeneracy among the parameters. 

With the fiducial model being the best-fitting value constrained by BAO data and \emph{Planck} $r_{\rm s}$ prior, we forecast the performance of upcoming DESI data, where the uncertainty on $H_0$ shrinks by a factor of $\sim 3.7$ (from 2.9 to 0.78), i.e., reduces by $\sim 73\% $ , which is quite encouraging. The improvement in the uncertainty of $H_0$ when the current data (SN$+$BAO$+$GLTD) are combined with LSST GLTD, DESI BAO, and LSST GLTD+DESI BAO are $\sim 10.8\%,\,37.8\%,$ and $ 38.3\%$, respectively, reaching $\sigma_{H_0} \approx 0.56$ uncertainty level. Our estimate of the improved $\sigma_{H_0} \approx 0.80$ with the inclusion of forecasted GLTD data alone, is in agreement with the analysis in \cite{Jee_2016}\footnote{A more recent analysis in \cite{Shiralilou19}, forecasts GLTD performance in an ideal scenario, which we do not compare with here. }. {Tentatively, the improved precision obtained with the future data (DESI+LSST) around the current best-fit model, would imply similar disagreements at the level of $\sim 2.2\sigma$ higher and $\sim 2.5\sigma$ lower value, than the \emph{Planck} $\Lambda$CDM and \citetalias{Riess_2018} $H_0$ derived values, respectively. This could imply a possibility for low-redshift ($0.1 \leq z \leq 2.5$) $H_0$ estimate that is in between the local ($z\leq 0.15$) and high-redshift CMB estimate.} As also shown in top panel of \Cref{mock}, the DESI BAO data contribute most to reducing the uncertainty. The LSST GLDT data are important as well, but they are overwhelmed by the BAO data constraining power when combined. Please note that we have not considered the additional distance information of $D_{\rm l}$ from the GLTD. According to \citet{Jee_2016}, including the $D_{\rm l}$ distance would improve the constraint significantly. Earlier forecast shows about 400 systems of robust measured time delay should be discovered by LSST \citep{Liao_2015}. We anticipate the future GLTD data will have a much better performance. Please note that the the fiducial cosmology to create the mock data sets being the best-fit of Log model to the current data, we do not study the contest between the GLTD and BAO data sets, but only forecast the precision of the joint constraint from the future low-redshift data. 

Finally, we repeat the exercise of replacing the $r_{\rm s}$ prior with the $r_{\rm s}h$ prior, as shown in the bottom panel of \Cref{mock}. The most significant improvements of constraints appear when including the mock BAO data using the $r_{\rm s}$ prior, while the LSST GLTD data provide only mild improvement. In contrast, when using the $r_{\rm s}h$ prior we find that the DESI BAO mock data, essentially do not provide any improvement to the constraints and that the major effect is driven by LSST GLTD data. This is simply representative of the fact that BAO data does not provide a constraint on $H_0$ unless $r_{\rm s}$ is known, either as an assumed prior or by inclusion of a dataset through which it is constrained. {Needless to say, the well-constrained higher value of $H_0 = 73.99\pm 0.80 \kmsMpc$ (Current data+LSST+DESI) is now accompanied by a lower value of $r_{\rm s} = 135.9 \pm 1.2\, \rm{Mpc}$, which is a $0.9\%$ constraint and a major improvement over the $3.3\%$ constraint from the current data.}

\section{Summary}
\label{sec:sum}
In the current work, we reconstruct the late-time expansion history of the universe in a cosmological-model-independent way, focusing on the Hubble constant $H_0$, using the latest SN Ia, BAO, and GLTD data, implementing four different parametric forms. A summary of our results is as follows:
\begin{itemize}[leftmargin=*]

\item Assuming the Gaussian prior on $r_{\rm s}$ from the high-redshift {\em{Planck}} estimate for $\Lambda$CDM, our deduced value of Hubble constant for the four models are more consistent with the \emph{Planck} $\Lambda$CDM, e.g., for the Log model, at $\sim 1.3\sigma$ using SN$+$BAO ($1.7\sigma$ using SN$+$BAO$+$GLTD) estimate than with the higher-valued local measurement at $\sim 2.5\sigma$ (2.3$\sigma$ using SN$+$BAO$+$GLTD). We find no preference among models having comparable values of DIC and assess the performance of the parametric models. 

\item With the updated data and also a curvature freedom (Log2 model), we reaffirm the conclusions of \citetalias{Lemos:2018smw}, that the Hubble tension possibly originates from the early universe. However, as the reconstructed $H(z)$, and hence $H_{\rm{0,rec}}$, is driven by the data (within the available freedom of the parametric models), conclusions remain to be verified with the more stringent future data.

\item Inclusion of GLTD data only mildly increases the best-fitting value of $H_{\rm{0,rec}}$, hardly improving uncertainty, due to the considerably lower constraining power of GLTD data and we assess mild disagreements among low-redshift data combinations. It is expected to yield possibly increased disagreements with the updated GLTD dataset in \cite{Wong19}.

\item Replacing the Gaussian $r_{\rm s}$ prior with the $r_{\rm s}h$ prior, we find a significant decrease of $\Delta\chi^2\sim 3$ of GLTD and a slight reduction for the BAO data. This further aids the argument that the early universe could be responsible for the Hubble tension, especially the comoving horizon $r_{\rm s}$. {A comparison of $r_{\rm s}h$ posteriors in these two cases, helps assess the allowed change in the angular scales constrained by CMB.}

\item More importantly, we anticipate the performance of future BAO and GLTD data from two upcoming experiments, DESI and LSST. When combined with the current data, we infer an improvement in uncertainty of $H_0$ by $\sim 10.8\%$ and $\sim37.8\%$, with GLTD and BAO data, respectively. {Combining these two future data will provide an improvement in precision by $\sim 38.3\%$, and might incite a need for agreement between local ($z\leq0.15$), low-redshift ($0.10\leq z \leq2.5$) and high-redshift (CMB) $H_0$ estimates, indicating moderate-level ($\ll 9\%$ of current difference) modifications to both the CMB and local $H_0$ estimates.}

\item {Replacing the $r_{\rm s}$ prior with the $r_{\rm s}h$ prior in the forecast analysis we find a value of $H_0$ consistent with \citetalias{Riess_2018}, and a lower value of $r_{\rm s} = 135.9 \pm 1.2\, \rm{Mpc}$, which is a $0.9\%$ constraint. This is a major improvement from the $3.3\%$ uncertainty, with the current data. }

\end{itemize}

{Implementing a multitude of contrasting analyses in a model-independent inverse distance ladder framework, we expect to find a strong degree of complementarity between BAO and GLTD data sets in the near future, which will provide tighter constraints on cosmological models, and also highlight much needed prospects to resolve the  $H_0$-tension and further important evidences to test physically motivated extensions to the $\Lambda$CDM model. }

\acknowledgments

\section*{Acknowledgments}
This work was supported by the National Natural Science Foundation of China under grants Nos. U1931202, 11633001, and 11690023. MV is supported by INFN INDARK PD51 grant and agreement ASI-INAF n.2017-14-H.0. BSH acknowledges financial support by ASI Grant No. 2016-24-H.0.

\bibliography{HZAPJ}

\end{document}